# NON-NEGATIVE MATRIX FACTORIZATION-CONVOLUTIONAL NEURAL NETWORK (NMF-CNN) FOR SOUND EVENT DETECTION

*Teck Kai Chan,[1,2]\*, Cheng Siong Chin[1]*　　*Ye Li[2]*

[1]Newcastle University Singapore
Faculty of Science, Agriculture, and Engineering
Singapore 599493
{t.k.chan2, cheng.chin}@newcastle.ac.uk

[2]Visenti Pte Ltd (A Brand of Xylem)
3A International Business Park
Singapore 609935
ye.li@xyleminc.com

**ABSTRACT**

The main scientific question of this year DCASE challenge, Task 4 - Sound Event Detection in Domestic Environments, is to investigate the types of data (strongly labeled synthetic data, weakly labeled data, unlabeled in domain data) required to achieve the best performing system. In this paper, we proposed a deep learning model that integrates Non-Negative Matrix Factorization (NMF) with Convolutional Neural Network (CNN). The key idea of such integration is to use NMF to provide an approximate strong label to the weakly labeled data. Such integration was able to achieve a higher event-based F1-score as compared to the baseline system (Evaluation Dataset: 30.39% vs. 23.7%, Validation Dataset: 31% vs. 25.8%). By comparing the validation results with other participants, the proposed system was ranked 8[th] among 19 teams (inclusive of the baseline system) in this year Task 4 challenge.

## 1. INTRODUCTION

The primary objective of a Sound Event Detection (SED) system is to identify the type of sound source present in an audio clip or recording and return the onset and offset of the identified source. Such a system has great potential in several domains, such as activity monitoring, environmental context understanding, and multimedia event detection [1], [2]. However, there are several challenges associated with SED in real-life scenarios.

Firstly, in real-life scenarios, different sound events can coincide [2]. Secondly, the presence of background noise could complicate the identification of sound event within a particular time frame [3]. This problem is further aggravated when the noise is the prominent sound source resulting in a low Signal to Noise Ratio (SNR). Thirdly, each event class is made up of different sound sources, e.g., a dog bark sound event can be produced from several breeds of dogs with different acoustic characteristics [1]. Finally, to achieve the best results, SED detection algorithm may require strongly labeled data where the occurrence of each event with its onset and offset are known with certainty during the model development phase.

While such data are useful, collecting them is often time-consuming, and sizes of such dataset are often limited to minutes or a few hours [3], [4]. In certain scenarios such as an approaching vehicle, the onset and offset time is ambiguous due to the fade in and fade out effect [5] and is subjective to the person labeling the event.

On the other hand, there exist a substantial amount of data known as the weakly labeled data where only the occurrence of an event is known without any offset or onset annotations. While it seems like the core information is missing, previous implementations proposed in the annual Detection and Classification of Acoustic Scenes and Events (DCASE) challenge that utilized only weakly labeled data had achieved a certain level of success [6]-[8]. Although a large number of different SED systems were proposed in the past, a majority of them were mainly based on Gaussian Mixture Model (GMM) [9], Hidden Markov Model (HMM) [10] or the use of dictionaries constructed using NMF [11-13]. However, due to the rising success of deep learning in other domains [14-17], deep learning for SED development is now a norm and has been shown to perform slightly better than established methods [1]. Riding on the success of deep learning, this paper proposes a deep learning model that integrates NMF and CNN which can provide an approximate strong label to the weakly labeled data. Results have shown that the proposed system achieved a much higher event based F1-score as compared to the baseline system (Evaluation Dataset: 30.39% vs. 23.7%, Validation Dataset: 31% vs. 25.8%) and by comparing the validation results with other participants, the proposed system was ranked 8[th] among 19 teams (inclusive of the baseline system) in this year Task 4 challenge.

\* This work was supported by the Economic Development Board-Industrial Postgraduate Programme (EDB-IPP) of Singapore under Grant BH180750 with Visenti Pte. Ltd.

## 2. RELATED WORK

In recent years, SED development has been overwhelmed with the use of deep learning algorithms particularly the use of CNN or Convolutional Recurrent Neural Network (CRNN). This phenomenon was also reflected in the 2018 and 2019 DCASE Task 4 challenge, where a large group of participants proposed the use of CRNN. As discussed in [1], CNN has the benefit of learning filters that are shifted in both time and frequency while Recurrent Neural Network (RNN) has a benefit of integrating information from the earlier time windows. Thus, a combined architecture has the potential to benefit from two different approaches that suggest its popularity.

The CRNN architecture proposed by Cakir et al. [1] first extracted features through multiple convolutional layers (with small filters spanning both time and frequency) and pooling in the frequency domain. The features were then fed to recurrent layers, whose features were used to obtain event activity probabilities through a feedforward fully connected layer. Evaluation over four different datasets had also shown that such a method has a better performance as compared to CNN, RNN and other established SED system. However, such a system would require a large amount of annotated data for training.

Lu [8] proposed the use of a Mean Teacher Convolution System that won the DCASE Task 4 challenge with an F1 score of 32.4%. In their system, context gating was used to emphasize the important parts of audio features in frames axis. Mean-Teacher semi-supervised method was then applied to exploit the availability of unlabeled data to average the model weights over training steps. Although this system won the 2018 challenge, there is still a large room for improvement.

## 3. SYSTEM OVERVIEW

### 3.1. Audio Processing

In this system, training inputs are mel-frequency scaled. This is because they can provide a reasonably good representation of the signal's spectral properties. At the same time, they also provide reasonably high inter-class variability to allow class discrimination by many different machine learning approaches [18].

In this paper, audio clips were first resampled to 32 kHz that were suggested to contain the most energies [19]. Moreover, segments containing higher frequency may not be useful for event detection in daily life [8].

A short-time fast Fourier transform with a Hanning window size of 1024 samples (32 ms) and a hop size of 500 samples (15.6 ms) was used to tabulate the spectrogram. After that, a mel filter bank of 64 and bandpass filter of 50 Hz to 14 kHz was applied to obtain the mel spectrogram. Finally, a logarithm operation was applied to obtain the log-mel spectrogram to use be used as input to the training model..

### 3.2. Non-Negative Matrix Factorization

The NMF popularized by Lee and Seung [20] is an effective method to decompose a non-negative $L \times N$ matrix $M$ into two non-negative matrices, $W$ and $H$ of sizes $L \times R$ and $R \times N$ respectively where $R$ is the number of components. The linear combination of $W$ and $H$ produces an approximated $M$ and can be represented as

$$M \approx WH \quad (1)$$

$W$ can be interpreted as the dictionary matrix and $H$ can be interpreted as the activation matrix. These two matrices can be randomly initialized and updated through the multiplicative rule given as [20] to produce an optimized set of $W$ and $H$. The updating procedure can be terminated when any further updating produces no improvement or when the difference of $M$ and $WH$ is below a user-defined threshold.

$$W \leftarrow W \otimes \frac{W^T \frac{M}{WH}}{W^T 1} \quad (2)$$

$$H \leftarrow H \otimes \frac{\frac{M}{WH} H^T}{1 H^T} \quad (3)$$

$W$ is commonly extracted on isolated events to form a dictionary and SED is performed by applying a threshold on the activation matrix obtained from the decomposition of the test data [12]. Since NMF only works on non-negative matrix, it was applied on the mel spectrogram prior to the logarithm operation. Thus, $M$ represents the mel spectrogram with $L$ as the number of mel bins and $N$ as the number of frames. In this paper, instead of consolidating $W$ to form the dictionary, we find the $H$ to indicate which frames of each audio clip are activated (above a pre-defined threshold) to label the weakly labeled data so that the weakly labeled data becomes an approximated strongly labeled data. If the clip contains multiple events, then those activated frames are deemed to contain all the sound events.

### 3.3. Convolutional Neural Network

The CNN used in this system is modified based on the one proposed in [19]. Kong et al. [19] proposed four different CNN with a different number of layers and pooling operators and found that the 9 layers CNN with max-pooling operator achieved the best performance. In this paper, we are interested in finding out whether with the inclusion of NMF, will a shallower CNN produce a comparable or even a better result. In this paper, a 5 layers CNN with max-pooling operator is proposed. In this architecture, the 5 layers consist of 1 input layer and 4 convolutional layers of kernel size 5 x 5 with a padding size of 2 x 2 and strides 1 x 1. This architecture is similar to Kong et al. [19] except for the kernel size and the number of layers as shown in Table 1.

Table 1. CNN architectures

| Proposed | Kong [19] |
|---|---|
| Input : log-mel spectrogram | |
| $\begin{pmatrix} 5 \times 5 @ 64 \\ BN, ReLU \end{pmatrix}$ | $\begin{pmatrix} 3 \times 3 @ 64 \\ BN, ReLU \end{pmatrix} \times 2$ |
| 2 x 2 Max Pooling | |
| $\begin{pmatrix} 5 \times 5 @ 128 \\ BN, ReLU \end{pmatrix}$ | $\begin{pmatrix} 3 \times 3 @ 128 \\ BN, ReLU \end{pmatrix} \times 2$ |
| 2 x 2 Max Pooling | |
| $\begin{pmatrix} 5 \times 5 @ 256 \\ BN, ReLU \end{pmatrix}$ | $\begin{pmatrix} 3 \times 3 @ 256 \\ BN, ReLU \end{pmatrix} \times 2$ |
| 2 x 2 Max Pooling | |
| $\begin{pmatrix} 5 \times 5 @ 512 \\ BN, ReLU \end{pmatrix}$ | $\begin{pmatrix} 3 \times 3 @ 512 \\ BN, ReLU \end{pmatrix} \times 2$ |

For both architectures, Binary Cross Entropy is adopted as the loss function which is similar to the loss function adopted in [19] given as

$$l_{BCE(p,y)} = \sum_{k=1}^{K}[y_k \ln(p_k) + (1-y_k)\ln(1-p_k)] \quad (4)$$

### 3.4. System Flow

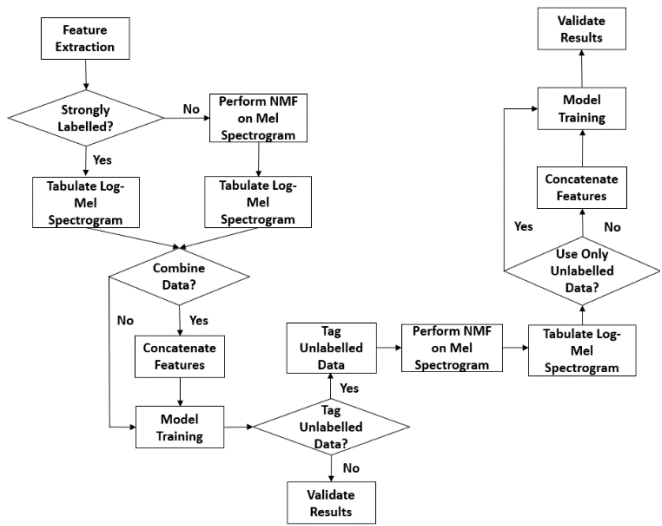

Fig. 1. Flowchart of proposed architecture

In this year DCASE challenge, Task 4 - Sound Event Detection In Domestic Environments, is specifically organized to investigate the types of data (strongly labeled synthetic data, weakly labeled data, unlabeled in domain data) required to achieve the best performing system. Therefore, the flow of the proposed system depends on the types of data used as seen in Fig. 1. While strongly labeled and weakly labeled data can be used readily, unlabeled data require a model to be trained in advance so that its content can be tagged and be used as training data. Example, if the user is keen to use all the data given, NMF will be applied to weakly labeled data to produce an approximated strongly labeled data. The log-mel spectrograms tabulated from both the approximated strongly labeled data and the actual strongly labeled data will be combined and used as input to the CNN. The model trained will be used to tag the events in the unlabeled data. Similar to weakly labeled data, NMF is applied to tagged unlabeled data prior to the calculation of log-mel spectrogram. These newly calculated log-mel spectrogram will be combined with previous calculated log-mel spectrograms and used as input to train a new model.

### 4. RESULTS AND DISCUSSION

Based on the proposed system flow, we tested the accuracy of our proposed architecture using the different combination of data on the given evaluation dataset that is a mixture of DCASE 2018 task 4 test set, and evaluation set consisting of 1168 audio clips with 4093 events.

Based on the results shown in Table 2, the model trained using both weakly labeled data and synthetic data achieved the highest accuracy as compared to using other combinations of data. It is surprising to find that strongly labeled synthetic data was not able to achieve higher accuracy than weakly labeled data. Whereas, a combination of data can increase the accuracy of the model.

On the other hand, results have shown that using only unlabeled in domain data or training a model with the inclusion of unlabeled in domain data labeled using different models, accuracy decreases. Furthermore, by comparing the proposed model results with Kong et al. [19] model and baseline model, it shows that although the proposed model can achieve a better event based F1 score, it has a lower segment based F1-score as compared to Kong et al. [19]. These two phenomenon could be due to the way how NMF was utilized. In this system, NMF was used to find H that indicates when the event was activated for the calculated H of certain frames were above a predefined threshold. However, if the clip contains multiple events, then NMF will indicate that those frames above a predefined threshold belong to all the events present in the audio. As such, it affected the quality of unlabeled data being labeled which resulted in a decrease in accuracy when unlabeled data is included and also resulted in a lower segment accuracy. Therefore, it may be worthwhile to investigate the use of source separation before the application of NMF.

The best four models (trained using C1, C3, C5, C7 as described in Table 2) were submitted to the challenge where

Detection and Classification of Acoustic Scenes and Events 201925-26 October 2019, New York, NY, USATable 2. F1-Score using different types of data, C1-Weakly Labeled Data, C2- Strongly Labeled Synthetic Data, C3- Weakly Labeled and Strongly Labeled Synthetic Data, C4- Unlabeled Data (labeled using model with F1 29.73%), C5- Weakly Labeled Data and Unlabeled Data (labeled using model with F1 29.73%), C6- Unlabeled Data (labeled using model with F1 30.39%), C7- Weakly Labeled, Strongly Labeled Synthetic Data and Unlabeled Data (labeled using model with F1 30.39%)

| F1-Score | Type of Data | C1 | C2 | C3 | C4 | C5 | C6 | C7 | Kong et al. [19] | Baseline [21] |
|---|---|---|---|---|---|---|---|---|---|---|
| Event Based F1 | Evaluation Dataset | 29.73% | 15.27% | **30.39%** | 25.47% | 27.2% | 26.64% | 27.84% | 24.1% | 23.7% |
| Segment Based F1 | | 55.79% | 43.59% | 57.66% | 45.88% | 48.52% | 47.13% | 50.92% | **63.0%** | 55.2% |
| Event Based F1 | Validation Dataset | 29.7% | - | **31.0%** | - | 26.9 | - | 27.7 | 22.3% | 25.8% |
| Segment Based F1 | | 55.6% | - | 58.2% | - | 48.7% | - | 50.5% | **59.4%** | 53.7% |

the validation dataset is made up of audio clips extracted from YouTube and Vimeo videos. The best performing was the model trained using C3 which achieved an F1-score of 31%. By comparing the validation results with other participants, the proposed system was ranked 8[th] among 19 teams (inclusive of the baseline system) in this year Task 4 challenge.

The system proposed by Lin and Wang [22] come in first place which achieved an accuracy of 42.7% while the system proposed by Yang et al. [23] took the last place which achieved an accuracy of 6.7%. On the other hand, the median accuracy for this challenge was at 29.25%.

While this system can be considered as an above-average system, there is still a large room of improvement as compared to the top 3 models which achieved F1-score of above 40%. One of the common features adopted by the top 3 models [22], [24], [25] was the use of mean teacher model which was also part of the winning model in 2018 [8] ([22] used a variant of mean teacher model called the professional teacher model). The idea of the mean teacher model was to average the model weights over training steps instead of label predictions and at the same time bringing the benefits of improved accuracy with fewer training labels [26] and this has become the new frontier in SED as seen in DCASE 2018 and 2019 challenge. However, it should be noted that virtual adversarial training as proposed by Agnone and Altaf [27] can be promising as well where it achieved an accuracy of 59.57% on the evaluation dataset although it only achieved an accuracy of 25% on the validation dataset. It was mentioned in [26] that both methods are compatible and their combination may produce better outcomes.

## 5. CONCLUSION

In this paper, a five layers CNN with the use of NMF was proposed for DCASE 2019 task 4. The proposed system was able to achieve a higher event based F1-score as compared to the baseline model. However, there is still room for improvement, particularly in the aspect of source separation that may very well helps in the accuracy of sound event detection. Future work may also consider the integration of mean teacher model virtual adversarial training which may produce an even better outcome.

## 6. ACKNOWLEDGMENT

We would like to thank the organizers for their technical support especially Nicolas Turpault, Romain Serizel and also Kong Qiuqiang from Surrey University for his prompt replies in regards to our questions on his system.

## 7. REFERENCES

[1] E. Cakır, G. Parascandolo, T. Heittola, H. Huttunen, and T. Virtanen, "Convolutional Recurrent Neural Networks for Polyphonic Sound Event Detection," *IEEE/ACM Trans Audio, Speech, an Language Process.*, vol. 25, no. 6, pp. 1291-1303, Jun. 2017.

[2] T. Hayashi, S. Watanabe, T. Toda, T. Hori, J. L. Roux, and K. Takeda, "Duration-Controlled LSTM for Polyphonic Sound Event Detection," *IEEE/ACM Trans Audio, Speech, an Language Process.*, vol. 25, no. 11, pp. 2059-2070, Nov. 2017.

[3] Q. Kong, Y. Xu, I. Sobieraj, W. Wang, and M. D. Plumbley "Sound Event Detection and Time–Frequency Segmentation from Weakly Labelled Data," *IEEE/ACM Trans Audio, Speech, an Language Process.*, vol. 27, no. 4, pp. 777-787, Apr. 2019.

[4] B. McFee, J. Salamon, and J. P. Bello, "Adaptive Pooling Operators for Weakly Labeled Sound Event Detection," *IEEE/ACM Trans Audio, Speech, an Language Process.*, vol. 26, no. 11, pp. 2180-2193, Apr. 2018.

[5] Q. Kong, Y. Xu, W. Wang, and M. D. Plumbley, "A Joint Separation-Classification Model For Sound Event Detection of Weakly Labelled Data," in *2018 IEEE Int. Conf. Acoustics, Speech and Signal Process. (ICASSP)*, Calgary, AB, Canada, Apr. 2018, pp. 321-325.

[6] S. Adavanne, G. Parascandolo, P. Pertila, T. Heittola, and T. Virtanen, "Sound Event Detection In Multichannel Audio Using Spatial and Harmonic Features," in *Detection and Classification of Acoustics Scenes and Events 2016 Workshop*, Budapest, Hungary, Sept. 2016, pp. 1-5.

[7] S. Adavanne, P. Pertila, and T. Virtanen, "Sound Event Detection Using Spatial Features and Convolutional

Recurrent Neural Network," in *Detection and Classification of Acoustics Scenes and Events 2017 Workshop*, Munich, Germany, Nov. 2017, pp. 1-5.

[8] J. Lu, "Mean Teacher Convolution System For DCASE 2018 Task 4," *Detection and Classification of Acoustics Scenes and Events 2018 Challenge*, Jul. 2018. [Online]. Available: http://dcase.community/documents/challenge2018/technical_reports/DCASE2018_Lu_19.pdf

[9] D. Su, X. Wu, L. Xu, "GMM-HMM acoustic model training by a two level procedure with Gaussian components determined by automatic model selection," in *2010 IEEE Int. Conf. Acoustics, Speech and Signal Process. (ICASSP)*, Dallas, TX, USA, Mar. 2010, pp. 4890-4893.

[10] A. Mesaros, T. Heittola, A. Eronen, and T. Virtanen "Acoustic Event Detection In Real Life," in *18th European Signal Process. Conf.*, Aalborg, Denmark, Aug. 2010, pp. 1267-1271.

[11] O. Dikmen, and A. Mesaros, "Sound Event Detection Using Non-Negative Dictionaries Learned From Annotated Overlapping Events," in *2013 IEEE Workshop Applications Signal Process. Audio Acoustics*, New Paltz, New York, Oct. 2013, pp. 1-4.

[12] V. Bisot, S. Essid, and G. Richard, "Overlapping Sound Event Detection With Supervised Nonnegative Matrix Factorization," in *2017 IEEE Int. Conf. Acoustics, Speech and Signal Process. (ICASSP)*, New Orleans, LA, USA, Mar. 2017, pp. 31-35.

[13] T. Komatsu, Y. Senda, and R. Kondo, "Acoustics Event Detection Based on Non-Negative Matrix Factorization With Mixtures of Local Dictionaries and Activation Aggregation," in *2016 IEEE Int. Conf. Acoustics, Speech and Signal Process. (ICASSP)*, Shanghai, China, Mar. 2016, pp. 2259-2263.

[14] Z. Md. Fadlullah, F. Tang, B. Mao, N. Kato, O. Akashi, T. Inoue, and K. Mizutani, "State-of-the-Art Deep Learning: Evolving Machine Intelligence Toward Tomorrow's Intelligent Network Traffic Control Systems," *IEEE Commun. Surveys Tutorials*, vol. 19, no. 4, pp. 2432-2455, 2017.

[15] Z. Liu, Z. Jia, C. Vong, S. Bu, J. Han, and X. Tang, "Capturing High-Discriminative Fault Features for Electronics-Rich Analog System via Deep Learning," *IEEE Trans. Indust. Inform.*, vol. 13, no. 3, pp. 1213-1226, Jun. 2017.

[16] M. He and D. He, "Deep Learning Based Approach for Bearing Fault Diagnosis," *IEEE Trans. Indust. Applications*, vol. 53, no. 3, pp. 3057-3065, Jun. 2017.

[17] T. Chan, K. Jia, S. Gao, J. Lu, Z. Zeng, and Y. Ma, "PCANet: A Simple Deep Learning Baseline for Image Classification?," *IEEE Trans. Image Process.*, vol. 24, no. 12, pp. 5017-5032, Dec. 2015.

[18] A. Mesaros, T. Heittola , E. Benetos, P. Foster, M. Lagrange, T. Virtanen, and M. D. Plumbley, "Detection and Classification of Acoustic Scenes and Events: Outcome of the DCASE 2016 Challenge," *IEEE/ACM Trans Audio, Speech, an Language Process.*, vol. 26, no. 2, pp. 379-393, Feb. 2018.

[19] Q. Kong, Y. Cao, T. Iqbal, Yong Xu, W. Wang, and M. D. Plumbley, "Cross-task learning for audio-tagging, sound event detection spatial localization: DCASE 2019 baseline systems," *arXiv preprint arXiv: 1904.03476*, pp. 1-5.

[20] D. D. Lee, and H. S. Seung, "Learning the parts of objects by non-negative matrix factorization," *Nature*, vol. 401, no. 6755, pp. 788-791, Oct. 1999.

[21] N. Turpault, R. Serizel, A. P. Shah, and J. Salamon, "*Sound event detection in domestic environments with weakly labeled data and soundscape synthesis,*" working paper or preprint, Jun. 2019. [Online]. Available: https://hal.inria.fr/hal-02160855.

[22] L. Lin and X. Wang, "Guided Learning Convolution System for DCASE 2019 Task 4," *Detection and Classification of Acoustics Scenes and Events 2019 Challenge*, Jun. 2019. [Online]. Available: http://dcase.community/documents/challenge2019/technical_reports/DCASE2019_Lin_25.pdf

[23] Q. Yang, J. Xia, and J. Wang, "Mean Teacher Model Based On CMRANN Network For Sound Event Detection," *Detection and Classification of Acoustics Scenes and Events 2019 Challenge*, Jun. 2019. [Online]. Available: http://dcase.community/documents/challenge2019/technical_reports/DCASE2019_Yang_18.pdf

[24] L. Delphin-Poulat and C. Plapous "Mean Teacher With Data Augmentation For DCASE 2019 Task 4," *Detection and Classification of Acoustics Scenes and Events 2019 Challenge*, Jun. 2019. [Online]. Available: http://dcase.community/documents/challenge2019/technical_reports/DCASE2019_Delphin_15.pdf

[25] Z. Shi, "Hodgepodge: Sound Event Detection Based On Ensemble of Semi-Supervised Learning Methods," *Detection and Classification of Acoustics Scenes and Events 2019 Challenge*, Jun. 2019. [Online]. Available: http://dcase.community/documents/challenge2019/technical_reports/DCASE2019_Shi_11.pdf

[26] A. Tarvainen and H. Valpola, "Mean teachers are better role models: Weight-averaged consistency targets improve semi-supervised deep learning results," in *31st Conf. Neural Information Processing Syst. (NIPS 2017)*, Long Beach, CA, USA, Dec. 2017, pp. 1-10.

[27] A. Agnone and U. Altaf, "Virtual Adversarial Training System For DCASE 2019 Task 4," *Detection and Classification of Acoustics Scenes and Events 2019 Challenge*, Jun. 2019. [Online]. Available: http://dcase.community/documents/challenge2019/technical_reports/DCASE2019_Agnone_78.pdf